\documentclass[12pt]{iopart}

\usepackage{graphics}
\usepackage{amssymb,color}
\usepackage{graphicx}
\usepackage{lipsum}
\usepackage{bbm}

\def\REV#1{\textcolor{black}{#1}}

\begin{document}

\title[$1/f$ noise and anomalous scaling in L\'evy on-off intermittency]{$1/f$ noise and anomalous scaling in Lévy noise-driven on-off intermittency  }

\author{Adrian van Kan$^1$ and François Pétrélis$^2$}   

\address{$^1$ Department of Physics, University of California at Berkeley, Berkeley, California 94720, USA\\
$^2$Laboratoire  de  Physique  de  l’Ecole  normale  sup\'erieure,  ENS,  Universit\'e  PSL, CNRS,  Sorbonne  Universit\'e,  Universit\'e  de  Paris,  F-75005  Paris,  France}
\vspace{10pt}
\begin{indented}
\item[]October 2022
\end{indented}

\begin{abstract}
On-off intermittency occurs in nonequilibrium physical systems close to bifurcation points, and is characterised by an aperiodic switching between a large-amplitude “on” state and a small-amplitude “off” state. L\'evy on-off intermittency is a recently introduced generalisation of on-off intermittency to multiplicative L\'evy noise, which depends on a stability parameter $\alpha$ and a skewness parameter $\beta$. Here, we derive two novel results on Lévy on-off intermittency by leveraging known exact results on the first-passage time statistics of L\'evy flights. First, we compute anomalous critical exponents \REV{explicitly} as a function of \REV{arbitrary} L\'evy noise parameters $(\alpha,\beta)$ \REV{for the first time, by a heuristic method, extending previous results}. The predictions are verified using numerical solutions of the fractional Fokker-Planck equation. Second, we derive the power spectrum $S(f)$ of L\'evy on-off intermittency, and show that it displays a power law $S(f)\propto f^\kappa$ at low frequencies $f$, where $\kappa\in (-1,0)$ depends on the noise parameters $\alpha,\beta$. An explicit expression for $\kappa$ is obtained in terms of $(\alpha,\beta)$. The predictions are verified using long time series realisations of Lévy on-off intermittency. Our findings help shed light on instabilities subject to non-equilibrium, power-law-distributed fluctuations, emphasizing that their properties can differ starkly from the case of Gaussian fluctuations. 
\end{abstract}

%
%
%
%
%

 \maketitle
\section{Introduction}
\label{intro}
Instabilities arise at parameter thresholds in many systems. Real physical systems are typically embedded in an uncontrolled noisy environment, \REV{with the noise deriving from high-dimensional chaotic dynamics}. The fluctuating properties of the environment affect the control parameter(s) of an instability, which leads to multiplicative noise. If this multiplicative noise is dominant over additive noise close to an instability threshold, the resulting behaviour is on-off intermittency, which is characterised by an aperiodic switching between a large-amplitude ``on" state and a small-amplitude ``off" (or "laminar") state, separated by some small threshold. It was extensively studied in the context of low-dimensional deterministic chaos and nonlinear maps \cite{fujisaka1985new,platt1993off,ott1994blowout,heagy1994characterization}, and has also been observed in numerous experimental setups ranging from electronic devices \cite{hammer1994experimental}, spin-wave instabilities \cite{rodelsperger1995off}, liquid crystals \cite{john1999off,vella2003off} and plasmas \cite{feng1998off} to multistable laser fibers \cite{huerta2008experimental}, sediment transport \cite{benavides2022impact}, human balancing motion \cite{cabrera2002off}, oscillator synchronisation \cite{yu1995off}, as well as blinking quantum dots in semiconductor nanocrystals \cite{margolin2005power,frantsuzov2008universal}, and measurements of earthquake occurence \cite{{bottiglieri2007off}}. On-off intermittency has also been observed in studies of in quasi-two-dimensional turbulence \cite{benavides2017critical,van2019condensates,van2021intermittency}, and magneto-hydroydnamic flows \cite{sweet2001blowout,alexakis2008effect,raynaud2013intermittency}. In addition, similar bursting behaviour is found in other contexts, including hydrodynamic \cite{sullivan1988nonperiodic,Knobloch1999,kumar2006critical} and neural systems \cite{hindmarsh1984model}. On-off intermittency has been investigated theoretically in the framework of nonlinear stochastic differential equations  \cite{aumaitre2005low,aumaitre2006effects,aumaitre2007noise} such as a pitchfork bifurcation with fluctuating growth rate,
\begin{equation}
    \frac{dX}{dt} = (f(t)+\mu)X - \gamma X^3, \label{eq:langevin}
\end{equation}
where $\mu\in \mathbb{R}$ and $f(t)$ is typically Gaussian white noise, with $\langle f(t) \rangle = 0$, $\langle f(t) f(t') \rangle = 2 \delta(t-t')$, in terms of the ensemble average $\langle \cdot \rangle$. Early studies of closely related models can be found in  \cite{horsthemke1976influence,yamada1986intermittency,fujisaka1986intermittency}. 
We can take $X$ to be positive without loss of generality, since (\ref{eq:langevin}) does \REV{not} allow sign changes. In the following, we adopt the Stratonovich interpretation of equation (\ref{eq:langevin}). A practical implication of this choice is that the rules of standard calculus apply to equation (\ref{eq:langevin}). For Gaussian white noise, the stationary probability distribution function (PDF) of the system is known to be of the form $p_{st}(x)= Nx ^{-1+\mu} e^{-\gamma x^2/2}$ with normalisation $N$, \REV{cf. \cite{horsthemke1976influence}}.  At small $\mu\geq 0$ the moments of $X$ scale as $\langle X^n \rangle\propto \mu^{c_n}$ with $c_n=1$ for all $n>0$, which is different from the deterministic ``mean-field" scaling $c_n = n/2$. This defines anomalous scaling, a well-known phenomenon in the context of continuous phase transitions \REV{(where noise is of thermal origin)} and critical phenomena \cite{kadanoff1967static,goldenfeld2018lectures}, as well as in turbulence \cite{eyink1994analogies,goldenfeld2017turbulence}. In addition to anomalous scaling, the result $c_n=1$ for all $n$ also implies multiscaling, \REV{which is defined by $c_n$ not being proportional to $n$}. Multiscaling occurs in a variety of contexts including turbulence \cite{frisch1993prediction}, finance \cite{di2007multi} and rainfall statistics \cite{gupta1990multiscaling}. \REV{In addition to its non-trivial scaling properties, the intermittent dynamics resulting from the multiplicative noise in equation (\ref{eq:langevin}) are reflected in the form of the power spectral density (PSD) of $X$. Denoting the two-time correlation function by $C(t)=\langle X(0) X(t)\rangle$, the Fourier transform of $C(t)$ defines the PSD of $X$, $S(f) = \int e^{ift} C(t) dt$, according to the Wiener-Khinchin theorem \cite{wiener1930generalized,khintchine1934korrelationstheorie}. This has been exploited to explain the $S(f)\propto f^{-1/2}$ range observed at low frequencies for small $\mu>0$, cf. \cite{manneville1980intermittency,petrelishabilitationa}. Such behaviour, namely the existence of a wide range in $\log(f)$, at small $f$, for which \REV{the PSD $S(f)$ is of power-law form with exponent smaller than $0$ and greater than $-2$}, is \REV{generically} called $\mathit{1/f}$ \textit{noise}, also known as {\it Flicker noise}, or \textit{pink noise}. It has been observed in a wide variety of systems, ranging from voltage and current fluctuations in vacuum tubes and transistors, where this behaviour was first recognised \cite{johnson1925schottky,hooge1981experimental,dutta1981low}, to astrophysical magnetic fields \cite{matthaeus1986low} and biological systems \cite{gilden19951}, climate \cite{fraedrich2003scaling}, turbulent flows \cite{dmitruk2007low,dmitruk2014magnetic,ravelet_chiffaudel_daviaud_2008}, reversing flows \cite{herault2015experimental,shukla2016statistical,pereira20191, dallas2020transitions}, traffic \cite{takayasu19931}, as well as music and speech \cite{voss19781,voss19751}, to name a few, and is also found in fractional renewal models \cite{watkins2017continuing}. In addition, $1/f$ noise has also been observed for Lévy flights in inhomogeneous environments \cite{kazakevivcius2014levy,kazakevivcius2015power}, but these studies did not consider any bifurcation points. }

\REV{While the above-described case of Gaussian noise has been studied in depth, non-Gaussian fluctuations arise in many systems. For example, out-of-equilibrium dynamics, such as turbulent fluid flows, typically exhibit non-Gaussian statistics, see e.g. \cite{she1991intermittency}, implying that instabilities developing on a turbulent background generally exhibit non-Gaussian growth rates, cf. \cite{van2021intermittency,alexakis2021symmetry}. Power-law-distributed fluctuations in particular are found in a variety of systems, including the human brain \cite{roberts2015heavy}, climate \cite{ditlevsen1999observation}, finance \cite{schinckus2013physicists} and beyond. An important example of random motion resulting from additive non-Gaussian noise is given by \textit{Lévy flights} (a term coined by Mandelbrot \cite{mandelbrot1983fractal}), which are driven by \textit{Lévy noise}.} Lévy noise follows a heavy-tailed $\alpha$-{\it stable} distribution that depends on a stability parameter $\alpha\in(0,2]$ and a skewness parameter $\beta\in [-1,1]$. Stable distributions come in different forms: the case $\alpha=2$ corresponds to the Gaussian distribution, while at $\alpha<2$ the distribution has power-law tails with exponent $-1-\alpha$. The main interest lies in the parameter regime $1<\alpha\leq 2$, where there is a finite mean, but an infinite variance. While the parameter regime $0<\alpha\leq 1$ is formally admissible, it is of little practical interest, since the noise distribution has a diverging mean in this case. The reason why Gaussian random variables are common in physics is their stability: by the central limit theorem \cite{feller2008introduction}, the Gaussian distribution constitutes an attractor in the space of PDFs with finite variance. Similarly, by the generalised central limit theorem \cite{gnedenko1954limit,uchaikin2011chance}, non-Gaussian $\alpha$-stable distributions constitute an attractor in the space of PDFs
whose variance does not exist. Stable distributions can be symmetric $(\beta=0)$ or asymmetric $(\beta\neq 0)$, giving rise to symmetric and asymmetric L\'evy flights. Lévy flights have since found numerous applications in many areas both in physics \cite{shlesinger1987levy,solomon1993observation,dubrulle1998truncated,metzler2000random,dubkov2008levy,del2005nondiffusive} and beyond, including climatology \cite{ditlevsen1999anomalous}, finance \cite{carr2003finite}, ecology \cite{reynolds2009levy} and human travel \cite{brockmann2006scaling}. 

\REV{Lévy statistics and on-off intermittency can be present in the same system. Examples include experiments of human balancing motion \cite{cabrera2002off,cabrera2004stick,cabrera2004human}, blinking quantum dots \cite{jung2002lineshape,margolin2005nonergodicity} and the intermittent growth of three-dimensional instabilities in quasi-two-dimensional turbulence \cite{van2021intermittency}. In a recent study \cite{vankan2021levy}, the problem of \textit{L\'evy on-off intermittency} was formally introduced as the case where $f(t)$ in equation (\ref{eq:langevin}) is L\'evy noise with $1<\alpha< 2$.} In this case, if $X(t)$ solves (\ref{eq:langevin}), then $\log(X(t))$ performs a L\'evy flight in an anharmonic potential. \REV{The asymptotics of the stationary PDF of $X$ were derived from the fractional Fokker-Planck equation associated with (\ref{eq:langevin}). However, an analytical solution for the full stationary PDF is only known in the Gaussian case ($\alpha=2)$. From the asymptotics of the stationary PDF, the moments $\langle X^n\rangle$ were computed heuristically in \cite{vankan2021levy}.} Anomalous scaling of the moments with the distance $\mu>0$ from the instability threshold was observed, with critical exponents $c_n$ that differ in general from the Gaussian case and depend on the stability and skewness parameters $\alpha$ and $\beta$ of the L\'evy noise. However, the explicit dependence of the critical exponents on $\alpha,\beta$ could only be computed for certain special cases in \cite{vankan2021levy}. Specifically, for all $-1<\beta<1$, the expression for the critical exponents obtained in \cite{vankan2021levy} contained a heuristic, numerically estimated constant. Therefore, it remains an open problem to determine the explicit dependence of the critical exponents on $\alpha,\beta$ at a theoretical level.  \REV{ In this paper we derive, for the first time, an explicit expression for the critical exponents in Lévy noise parameters with arbitrary parameters $\alpha,\beta$, using heuristic arguments. Moreover, although the power spectral density in on-off intermittency with Lévy statistics has been experimentally measured for human balancing motion, where a low-frequency exponent close to $-1/2$ was found \cite{cabrera2002off}, no theoretical results exist for the PSD of Lévy on-off intermittency, and the dependence of the noise parameters remains unknown. Here, we present a heuristic derivation of the low-frequency PSD in Lévy on-off intermittency. Both derivations will be explicated later on in the text.}

In addition to critical scaling, another important characteristic of on-off intermittency is given by the statistics of the duration $T_{off}$ of laminar phases. These have received much attention, in particular since they are rather easily accessible numerically \cite{platt1993off,ott1994blowout,heagy1994characterization} and in experiments \cite{hammer1994experimental,rodelsperger1995off,john1999off,vella2003off,feng1998off,huerta2008experimental, cabrera2002off,benavides2022impact,margolin2005power,frantsuzov2008universal,bottiglieri2007off}.  In many studies, $T_{off}$ is found to follow a PDF with a power-law tail $p(T_{off})\propto (T_{off})^{m}$, with $m=-3/2$. The value of the exponent has been explained in terms of first-passage time statistics: on a logarithmic scale, the linear dynamics in the laminar phase can be mapped onto a random walk on the negative half line, so that the duration of laminar periods corresponds to the first-passage time through the origin of the random walk. For symmetric random walks, this quantity is known to follow a PDF with a power-law tail whose exponent is $-3/2$ \cite{redner2001guide}. According to the Sparre Andersen theorem \cite{andersen1954fluctuations,klafter2011first}, 
 this result holds for any symmetric step size distribution, as long as steps are independent, including symmetric L\'evy flights\REV{, for which $\beta=0$,} and even in the presence of finite spatio-temporal correlations \cite{artuso2014sparre}. 
Despite the large body of research corroborating the scenario leading to the exponent $m=-3/2$, some studies on blinking quantum dots, bubble dynamics and other systems \cite{kuno2001off,kuno2003modeling,divoux2009intermittent,bertin2012off} find a different behaviour. There, the duration of laminar phases also follows a power-law distribution, but with exponent $m\neq -3/2$ varying between $-1$ and $-2$. Similarly, Manneville \cite{manneville1980intermittency} finds $m=-2$ for a chaotic discrete map. For Lévy flights, there exist exact results for the  asymptotics of the first-passage time distribution. The distribution features a power-law tail with exponent \REV{$m=m(\alpha,\beta) \in(-1,-2)$}, whose dependence on ($\alpha,\beta$) is known explicitly. A summary and derivation of these results is given in \cite{padash2019first}. The goal of the present paper is to leverage these exact first-passage time results to better understand two aspects of L\'evy on-off intermittency: its anomalous critical exponents close to the threshold of instability, and its power spectral density (PSD).

For the case of Gaussian white noise $f(t)$ in equation (\ref{eq:langevin}), where the critical exponents can be calculated directly from the known stationary PDF, an alternative derivation was presented in the work of Auma\^itre et al. \cite{aumaitre2006effects}, where a heuristic argument based on the knowledge of $p(T_{off})$ and simple properties of the on-phases leads to the same result. In the present study, we first generalise the argument given by Aumaître et al. to L\'evy on-off intermittency, where the stationary PDF is not fully known. \REV{We thus derive, for the first time, explicit expressions for the critical exponents valid for arbitrary noise parameters $\alpha,\beta$.} First-passage time statistics are also known to be linked to the two-time correlation function $C(t) = \langle x(t) x(0) \rangle$ in on-off intermittency, \REV{as described in \cite{manneville1980intermittency,petrelishabilitationa}. In the second part of this paper, we generalise these arguments to L\'evy on-off intermittency to show that it displays $1/f$ noise with a spectral power-law range $S(f)\propto f^{\kappa}$ whose exponent $\kappa\in(-1,0)$} \REV{is computed explicitly for the first time, and shown to depend on the noise parameters.}

The remainder of this paper is structured as follows. In section \ref{sec:theoretical_background}, we describe the theoretical background of this study. Next, in section \ref{sec:crit_exp} we present a derivation of the critical exponents in L\'evy on-off intermittency, comparing the results to the findings of \cite{vankan2021levy} and additional numerical solutions of the fractional Fokker-Planck equation associated with equation (\ref{eq:langevin}). In section \ref{sec:flicker}, we present a spectral analysis of Lévy on-off intermittency. We describe the arguments relating first-passage time distributions to $1/f$ noise and again verify the predictions numerically. Finally, in section \ref{sec:conclusions}, we discuss our results and conclude.
\section{Theoretical background}
\label{sec:theoretical_background}
In this section we define stable distributions, recall results on L\'evy on-off intermittency, and introduce relevant properties of L\'evy flight first-passage time PDFs.

\subsection{Definition of $\alpha$-stable distributions}
For parameters $\alpha\in (0,2],\beta\in [-1,1]$, we denote the alpha-stable PDF for a random variable $Y$ by $\wp_{\alpha,\beta}(y)$. It is defined by its characteristic function (i.e. Fourier transform),
\begin{equation}
    \varphi_{\alpha,\beta}(k) = \exp\Bigg\lbrace - |k|^\alpha [1-i\beta\mathrm{sgn}(k) \Phi(k)] \Bigg\rbrace, \label{eq:def_stab_cf}
\end{equation}
 with
\begin{equation}
    \Phi(k) = 
     \tan\left( \frac{\pi \alpha}{2} \right) \hspace{0.25cm} \textrm{ for } \alpha \neq 1 ,\hspace{1.5cm} \Phi(k) = - \frac{2}{\pi} \log(|k|) \hspace{0.25cm} \textrm{ for }\alpha =1, \hspace{1.5cm} 
    \end{equation}
see \cite{uchaikin2011chance}. We note that (\ref{eq:def_stab_cf}) is not the most general form possible: there may be a scale parameter in the exponential, which we set equal to one. One refers to $\alpha$ as the stability parameter. For $\alpha=2$, one recovers the Gaussian distribution, independently of the skewness parameter $\beta$. For $\alpha<2$, $\beta$ controls the asymmetry of the distribution, with perfect symmetry at $\beta=0$, and the strongest asymmetry at $|\beta|=1$. For $|\beta|<1$, the stable PDF has two power-law tails at $y\to\pm \infty$, where $\wp_{\alpha,\beta}(y)\sim \lbrace1+\beta \mathrm{sign}(y)\rbrace|y|^{-1-\alpha}$. For $\beta=\pm1$, the prefactor vanishes in one limit, and the asymptote at $y\to -\beta \infty$ changes from power-law to exponential decay. In the following, we restrict our attention to $1<\alpha<2$, since on-off intermittency only occurs in this parameter range (and in the Gaussian case $\alpha=2$), \REV{as the mean $\langle Y \rangle$, with $Y$ drawn from $\wp_{\alpha,\beta}(y)$, only exists for $\alpha>1$. It is important to note that the definition of the stable distributions implies $\langle Y \rangle=0$, whether its underlying distribution $\wp_{\alpha,\beta}(y)$ is symmetric or not. By contrast, the most probable value of $Y$, corresponding to the maximum of $\wp_{\alpha,\beta}(y)$, is only equal to zero for $\beta=0$, and is non-zero for $\beta\neq 0$.}

\subsection{Relevant results pertaining to L\'evy on-off intermittency}
\label{sec:relevant_levy_on_off}
\REV{Here we recall some important results obtained in \cite{vankan2021levy}.} As in \cite{vankan2021levy}, we consider the  Langevin equation (\ref{eq:langevin}) with $f(t)$ being white Lévy noise, which follows an $\alpha$-stable distribution. More  precisely, for a given  time step $dt$, we let $f(t)dt = dt^{1/\alpha}F(t),$ where $F(t)$ obeys the alpha-stable PDF $\wp_{\alpha,\beta}(F)$, defined by (\ref{eq:def_stab_cf}), and is drawn independently at every time step $t$, cf. also \cite{dubkov2008levy}. For these dynamics, critical exponents were computed from the fractional Fokker-Planck equation associated with (\ref{eq:langevin}) in  \cite{vankan2021levy}. \REV{For $1<\alpha\leq 2$ (i.e. when the mean of $f(t)$ exists), equation (\ref{eq:langevin}) implies
\begin{equation}
\frac{d \langle \log(X) \rangle}{dt} = \mu +\underbrace{\langle f(t)\rangle}_{=0} - \gamma \langle X^2 \rangle. \label{eq:exact_identity}
\end{equation}
For $\mu >0$, the system reaches a steady state where $d\langle \log(X)\rangle /dt =0$, implying
\begin{equation}
\langle X^2\rangle= \mu/\gamma, \label{eq:x2_exact}
\end{equation} which is an exact relation showing that the second-order moment exists, and that the associated critical exponent $c_2=1$ for all $\alpha,\beta$. If, on the other hand, $\mu<0$, then no such steady state exists, and the right-hand side of equation (\ref{eq:exact_identity}) is strictly negative. This indicates that the threshold of the instability is set by $\mu+\langle f(t)\rangle =\mu=0$, independently of noise parameters, including the case where $f(t)$ is asymmetric ($\beta\neq 0$).}  
\REV{The existence of moments can be deduced from the large-$X$ asymptotics of the stationary PDF, which for nonlinearity of order $s$ ($s=3$ in equation (\ref{eq:langevin})), and $\beta>-1$, is given by 
\begin{equation}
p_{st}(x) \sim C \frac{(1+\beta)}{\gamma} x^{-s} \log^{-\alpha}(x) \hspace{1.5cm} \textrm{as}\hspace{1.5cm} x \to \infty,\label{eq:asymp_pdf_beta_gt_m1}
\end{equation}
with $C=\sin(\pi \alpha/2) \Gamma(\alpha)/\pi$. This asymptotic result, a straightforward generalisation of the cubic case considered in \cite{vankan2021levy}, implies that for a nonlinearity of order $s$ the first $s-1$ moments exist. Furthermore the moment of order $s-1$ is special, in that it is convergent only due to the logarithmic factor in the PDF (provided $\alpha >1$), and therefore converges slowly at large $X$. The case $\beta=-1$ is an exception, where the PDF decays exponentially at large $X$, and therefore $\langle X^n\rangle $ exists for all $n$, independently of the order of the nonlinearity, and all critical exponents are equal to unity, as in the Gaussian case. Moments of any order $n>s-1$, where $s$ is the order of nonlinearity, are found to diverge.}  
For $-1<\beta<1$, the first moment was predicted by \cite{vankan2021levy} to scale as $\langle X\rangle \propto \mu^{c_1}$, with 
\begin{equation}
c_1= (1-\nu)/(\alpha-1), \label{eq:cn_beta0}
\end{equation}
in terms of a $\mu$-independent parameter $\nu$, which could only be determined numerically, with significant uncertainties, and whose full dependence on $\alpha$,$\beta$ remains unknown. For $\alpha=1.5$, $\beta=0$, it was found numerically that $\nu\approx 0.25$. \REV{Here, we will compute $\nu$ explicitly as a function of general noise parameters $\alpha,\beta$}.

Another important result derived in \cite{vankan2021levy} from the fractional Fokker-Planck equation associated with equation (\ref{eq:langevin}) is that, with two exceptions, Lévy on-off intermittency occurs for any positive value of the control parameter $\mu$. \REV{This is due to the fact that the stationary PDF of $X$ is asymptotically given by
\begin{equation}
p_{st}(x)\sim C(1-\beta) (\mu x)^{-1} \log^{-\alpha}(1/x) \hspace{1.5cm} \textrm{for} \hspace{1.5cm} 0<x\ll 1,
\end{equation}
with $C$ as in equation (\ref{eq:asymp_pdf_beta_gt_m1}), such that the most probable state is always $x=0$.} An exception arises for $\beta=1$, where the above asymptotic relation breaks down, and the stationary PDF is of the form $p_{st}(x)\propto x^{-1 + A_\alpha(\mu)}$, where $A_\alpha(\mu)>0$ increases monotonically with $\mu$, and therefore on-off intermittency ceases at $\mu=\mu^*$, where $A_\alpha(\mu^*)=1$. This is analogous to the case of Gaussian noise $(\alpha=2$), where $A_2(\mu)=\mu$, and thus on-off intermittency is also only observed in a finite interval of the control parameter $\mu$ there.
\begin{figure}
\centering
\includegraphics[width=0.5\textwidth]{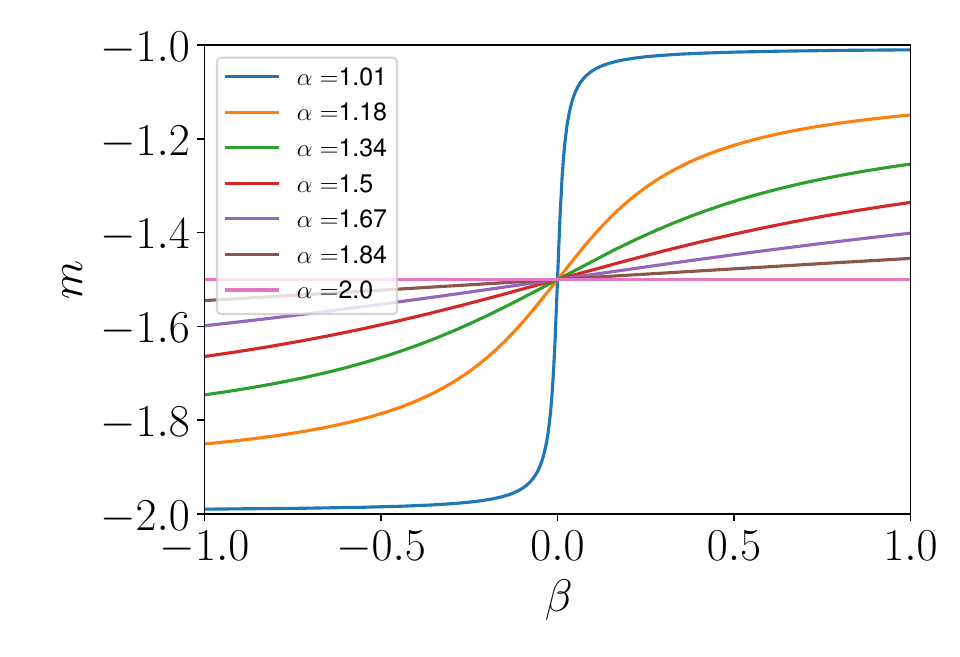}
\caption{Exponent $m$ defined in equation (\ref{eq:m_alpha_beta}), shown versus $\beta$ for different $1<\alpha\leq 2$. The value of $m$ increases monotonically with $\beta$, and lies in the interval $(-2,-1)$.  }
\label{fig:m_al_beta}
\end{figure}
\subsection{First-passage time distributions of L\'evy flights}
Due to their importance in many applications, first-passage problems have received much attention in both standard Brownian motion \cite{redner2001guide} and L\'evy flights \cite{padash2019first}. Consider equation (\ref{eq:langevin}) with $\mu>0$ and $\gamma=0$, restricted in terms of $Y=\log(X)$ to the negative semi-infinite half line with an absorbing boundary at $Y=0$. Choose an initial condition $y=-d<0$. For $\alpha=2$, i.e. standard Brownian motion with a drift, the first-passage time (FPT) $\tau$ follows the so-called L\'evy distribution \cite{redner2001guide} (a special case of stable distributions with $\alpha=1/2$, $\beta=1$),
\begin{equation}
\mathcal{P}(\tau) = \frac{d}{(4\pi \tau)^{3/2}} \exp\left(-\frac{(d-\mu \tau)^2}{4\tau}\right). \label{eq:FPT_dist_BM}
\end{equation}
At small times $\tau \ll d/\mu$, $\mathcal{P}(\tau)$ vanishes, since it takes a finite time to reach the absorbing boundary. For intermediate times $\tau$ with $d \ll \mu \tau \ll 2\sqrt{\tau}$, one finds $\mathcal{P}(\tau)\propto \tau^{-3/2}$. The power law is eventually cut off by the exponential factor at $\tau= t_{c}(\mu)\propto \mu^{-2}$. The cut-off time is set by the cross-over between the diffusive motion at early times $t$, where the typical displacement grows as $\sqrt{t}$, and the ballistic motion $y= \mu t$ at late times. After the time $t_c(\mu)$, the trajectory has almost certainly reached the absorbing boundary at $y=0$ due to the drift and hence $P(\tau)$ vanishes again.

For Lévy flights, the mean first passage time is not known in full for arbitrary parameter values $\alpha,\beta$. However, for vanishing drift, $\mu=0$, and $1<\alpha<2$, $\beta\in[-1,1]$, the L\'evy flight first-passage time distribution has been shown \cite{padash2019first} to be asymptotically proportional to 
\begin{equation}
\mathcal{P}(\tau)\propto \tau^m \phantom{xxx}\mathrm{ with }\phantom{xxx} m(\alpha,\beta) = -3/2-(\alpha \pi)^{-1} \arctan[\beta \tan(\pi\alpha/2)].
\label{eq:m_alpha_beta}
\end{equation}
In particular, for the case of symmetric noise ($\beta=0$), this reduces to $m=-3/2$ as in the Gaussian case, in agreement with the Sparre-Andersen theorem. For $\beta=1$, one finds $m=-2+1/\alpha$, and for $\beta=-1$ one gets $m=-1-1/\alpha$. As $\alpha$ varies from $1$ to $2$ and $\beta$ from $-1$ to $1$, the exponent $m$ varies continuously between $-1$ and $-2$. Figure \ref{fig:m_al_beta} illustrates the dependence of $m$ on $\alpha,\beta$. 

\begin{figure}
\centering
\includegraphics[width=8.6cm]{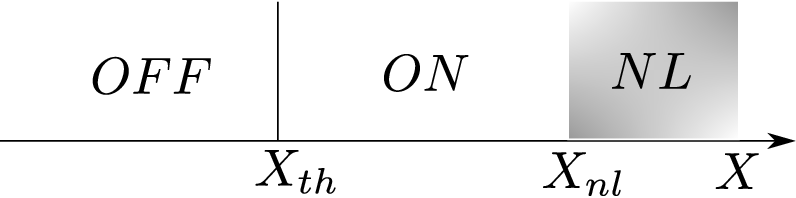}
\caption{Sketch of heuristic model for equation (\ref{eq:langevin}). An arbitrary small threshold $X_{th}$ \REV{separates the finite-size on domain from the semi-infinite off domain.} The effect of nonlinearity (NL) is modeled by a wall at $X=X_{nl}$. }
\label{fig:sketch}
\end{figure}
\section{Critical exponents}
\label{sec:crit_exp}
Here we give a simple heuristic argument connecting the statistics of L\'evy flight first-passage times to the anomalous scaling of moments in L\'evy on-off intermittency. We first reproduce the argument of Aumaître et al. given in \cite{aumaitre2006effects} for Gaussian noise, then we go on to generalise it to L\'evy noise. For the remainder of this paper, we will take $\gamma=1$ in equation (\ref{eq:langevin}).

\subsection{The Gaussian case}
We first consider Gaussian noise, i.e. $\alpha=2$. \REV{In the exact stationary PDF of $X$, $p_{st}(x) = N x^{-1+\mu} e^{-\gamma x^2/2}$, the only impact of the nonlinearity is to provide an exponential cut-off at large $x$. This motivates a study of the situation depicted in figure \ref{fig:sketch}, where the suppression of large amplitudes by the nonlinear is modeled as a reflective wall positioned at a large $X=X_{nl}$, and an arbitary, threshold $X_{th}$ is defined to separate the semi-infinite off domain from the finite-size on domain.} In the off domain, $\log(X)$ performs simple Brownian motion with drift $\mu$, which is assumed positive. While the off domain is a semi-infinite interval, the on domain is finite. Starting from within the off domain, one can compute the mean first-passage time through $X=X_{th}$ using (\ref{eq:FPT_dist_BM}). One finds 
\begin{equation}
\langle T_{off} \rangle \propto \sqrt{t_c(\mu)} \label{eq:Toff_Gaussian}
\end{equation}
with $t_c(\mu)\propto \mu^{-2}$. Once the trajectory crosses the threshold $X=X_{th}$, it remains in the on phase until it exits by diffusion and nonlinear damping (which compete against the positive drift), and the process repeats. 
We denote by $T_{on,tot}$ the total time spent in the on state after a long simulation time $T$. Then the fraction of time spent in the on phase over the full simulation time, $T_{on,tot}/T$, is given by the average duration $\langle T_{on}\rangle$ of an on phase, multiplied by the frequency of occurence of on phases. The latter is approximately equal to $1/\langle T_{off} \rangle$, which is known from equation (\ref{eq:Toff_Gaussian}). Aumaître et al. argue that $\langle T_{on}\rangle$ tends to a finite, $\mu$-independent constant as $\mu \to 0^+$. Hence,
\begin{equation}
\frac{T_{on,tot}}{T}\propto \mu.
\end{equation}  
We denote by $\langle X^n\rangle_{on}$ the value of $X^n$ averaged over on phases. For small $\mu>0$, $\langle X^n\rangle_{on}$ becomes independent of $\mu$. This can be understood using the analogy with a random walk in a finite interval delimited by a wall on one side and an absorbing boundary on the other: \REV{typical trajectories of the random walk \REV{to exit the finite on domain} are dominated by diffusion for small $\mu$, and therefore do not depend on $\mu$ in the small-$\mu$ limit.} This finally implies that the moments scale as
\begin{equation}
\langle X^n \rangle \propto \langle X^n\rangle_{on} \times T_{on,tot}/T \propto \mu
\end{equation}
for all $n>0$, which is precisely what is found when computing the moments explicitly from the stationary PDF.
\begin{figure}[h]
\centering
\includegraphics[width=0.49\textwidth]{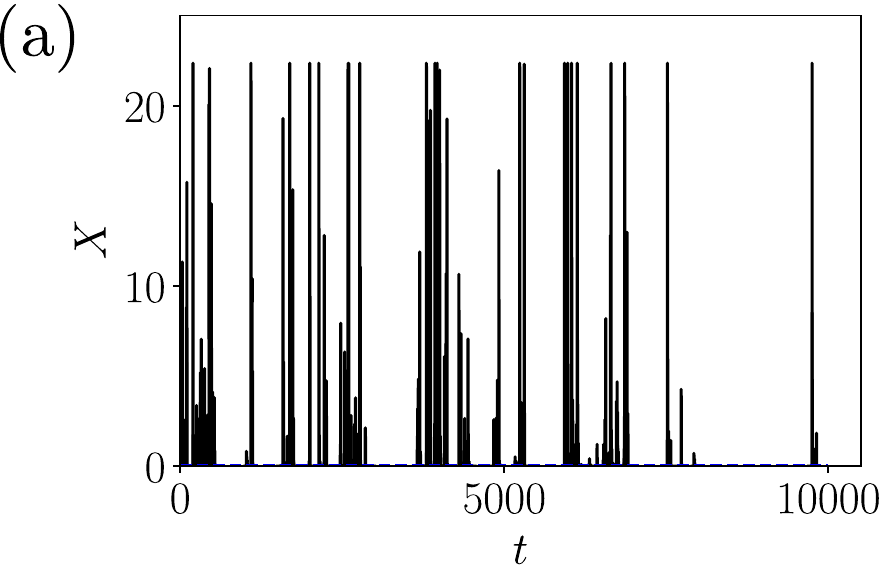}
\includegraphics[width=0.49\textwidth]{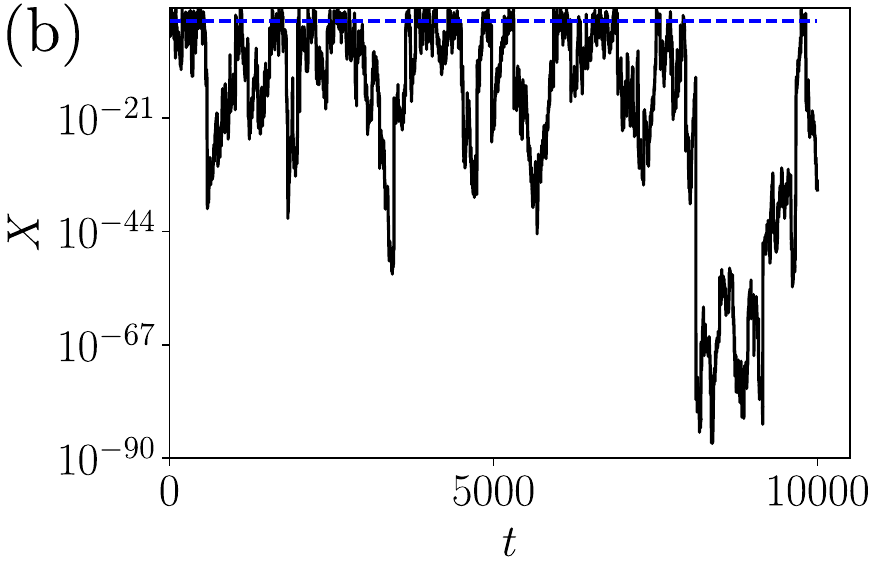}\\
\hspace{0.011\textwidth}\includegraphics[width=0.49\textwidth]{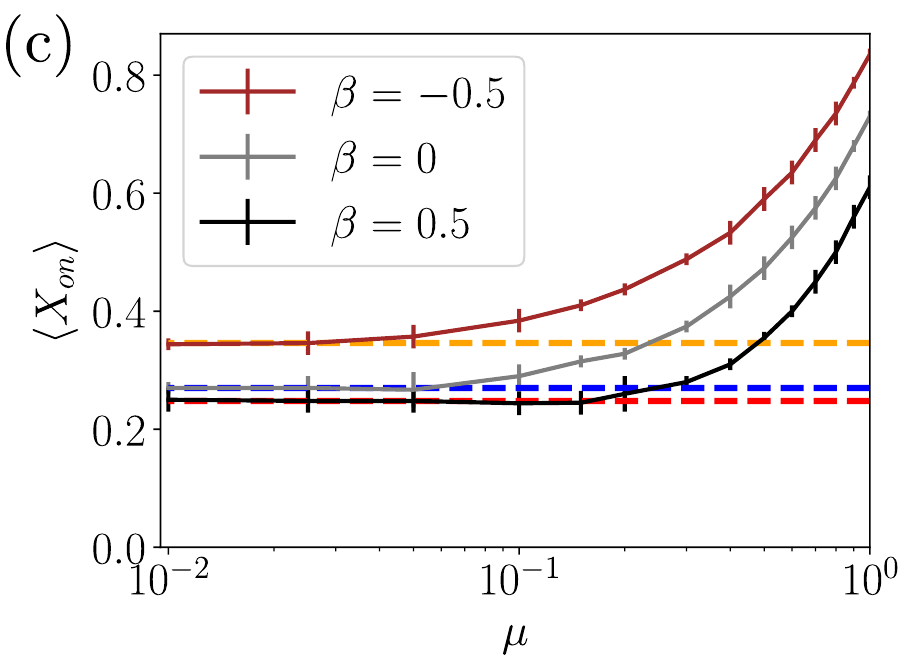}
\includegraphics[width=0.49\textwidth]{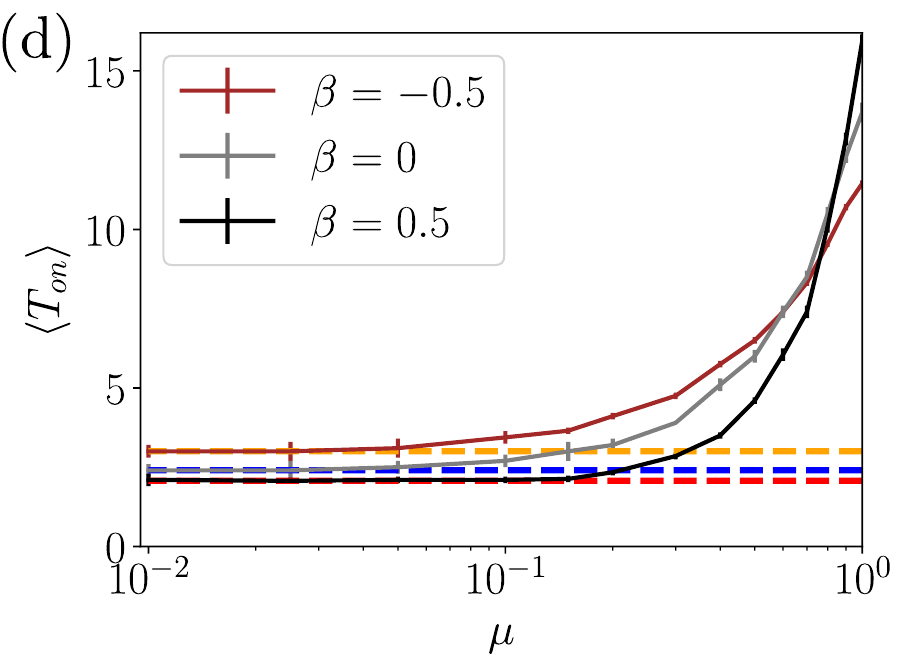}
\caption{Panel a): Time series displaying on-off intermittency at $\alpha=1.5$, $\beta=0$, $\mu=0.1$ and $\gamma=1$, generated using the formal solution of the Langevin equation (\ref{eq:langevin}) given in the appendix of \cite{vankan2021levy}, with the dashed blue line indicating $X_{th}=0.001$. Panel b) Same time series with a logarithmic $y$-axis. Panel c): Average value of $X$ during on phases, computed based on similar time series as in panel a) with time step $dt=0.01$, at $\alpha=1.5$, $\beta=-0.5,0,0.5$ for nonlinear coefficient $\gamma=1$, as a function of $\mu$. Panel d): same as panel c for average duration of on phases.}
\label{fig:Xon_Ton_vs_mu}
\end{figure}
\subsection{The Lévy case}
We now generalise the above argument to L\'evy on-off intermittency, for which a typical time series is shown in figure \ref{fig:Xon_Ton_vs_mu}a. All time series of $X$ computed in this work are generated using the formal solution of equation (\ref{eq:langevin}) given in \cite{vankan2021levy}. We primarily focus on the case $|\beta|<1$, where the noise follows a distribution with power-law tails at both positive and negative values, that can be symmetric ($\beta=0$) or asymmetric ($\beta\neq 0$). Consider again the heuristic model depicted in figure \ref{fig:sketch}, with a sharp cut-off by nonlinearity at $X=X_{nl}$ as a simplified description of equation (\ref{eq:langevin}). We hasten to add that, for Lévy flights, as discussed in \cite{dybiec2004resonant}, the implementation of reflecting boundary conditions is non-trivial due to the possibility of leapovers \cite{koren2007first,koren2007leapover}, which make it possible for a trajectory to pass a point without hitting it \cite{palyulin2019first}. Also, by contrast with the Gaussian case, the order $s$ of nonlinearity ($s=3$ in eq. (\ref{eq:langevin})) impacts the moments non-trivially: the number of finite integer-order moments of $X$ in stationary state is equal to $s-1$, except in the case $\beta=-1$, as discussed in the introduction. At best, one can hope that the model depicted in fig. \ref{fig:sketch} may reproduce the existing, finite moments of order $n<s-1$ correctly. \REV{The moment of order $s-1$, which is fixed by the exact identity (\ref{eq:x2_exact}), derives from a slowly converging integral at $X\to \infty$, as explained in section \ref{sec:relevant_levy_on_off}, and therefore cannot be captured by the present argument (there is no sharp cut-off by the nonlinearity in that case).} Notwithstanding these caveats, we proceed on the modelling assumption and verify a posteriori that the predictions are consistent with the known results of \cite{vankan2021levy} and additional simulations. Using the asymptotics given in eq. (\ref{eq:m_alpha_beta}), we can deduce that the mean time spent in the off state scales as
\begin{equation}
\langle T_{off} \rangle \propto t_c^{m(\alpha,\beta)+2}, \label{eq:Toff_mean_Levy}
\end{equation}
where $t_c$ is again a cut-off time. While in the Gaussian case {$t_c(\mu)$} is known from the full FPT distribution (\ref{eq:FPT_dist_BM}), this is not the case for Lévy flights, since the result in equation (\ref{eq:m_alpha_beta}) does not include a finite drift. However, {$t_c(\mu)$} may be determined as the cross-over time between drift and L\'evy flight superdiffusive motion. The typical distance travelled superdiffusively in a L\'evy flight after time $t$ is proportional to $t^{1/\alpha}$, see \cite{dubkov2008levy}. Since we consider $1<\alpha<2$, such that $1/\alpha <1$ the drift $\mu t$ is initially small compared to superdiffusion, but eventually dominates after a finite cross-over time. Its value is found by balancing $t^{1/\alpha}$ with the drift $\mu t$ (we consider $\mu>0$), giving
\begin{equation}
t_c \propto \mu^{\alpha/(1-\alpha)}. \label{eq:def_tc}
\end{equation} 
This can now be combined with equation (\ref{eq:Toff_mean_Levy}) to give the dependence of the mean first passage time on $\alpha,\beta$ and $\mu$. In addition, as shown in figure \ref{fig:Xon_Ton_vs_mu}d, the mean duration of on phases $\langle T_{on}\rangle$ tends to a $\mu$-independent constant as $\mu\to 0^+$ for $|\beta|<1$, like in the Gaussian case. Hence, the total time $T_{on,tot}$ spent in the on state for a time series of length $T$ satisfies 
\begin{equation}
\frac{T_{on,tot}}{T} \approx \langle T_{on} \rangle / \langle T_{off} \rangle \propto \mu^{\alpha(m(\alpha,\beta)+2)/(\alpha-1)}
\end{equation}
at small $\mu>0$.

\begin{figure}
\centering
\includegraphics[width=0.46\textwidth,height=0.37\textwidth]{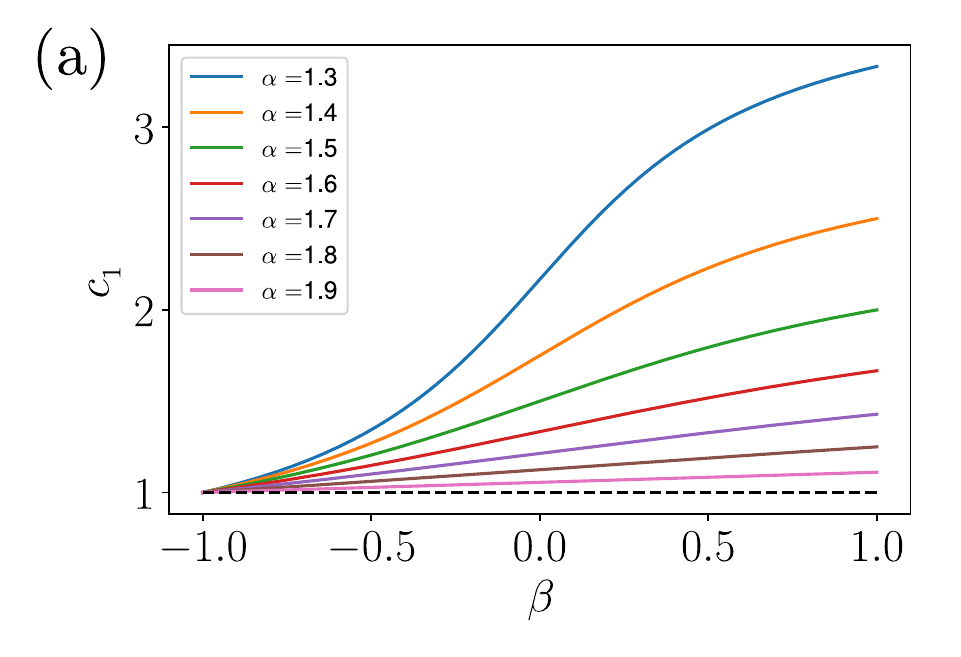}
\includegraphics[width=0.47\textwidth,height=0.38\textwidth]{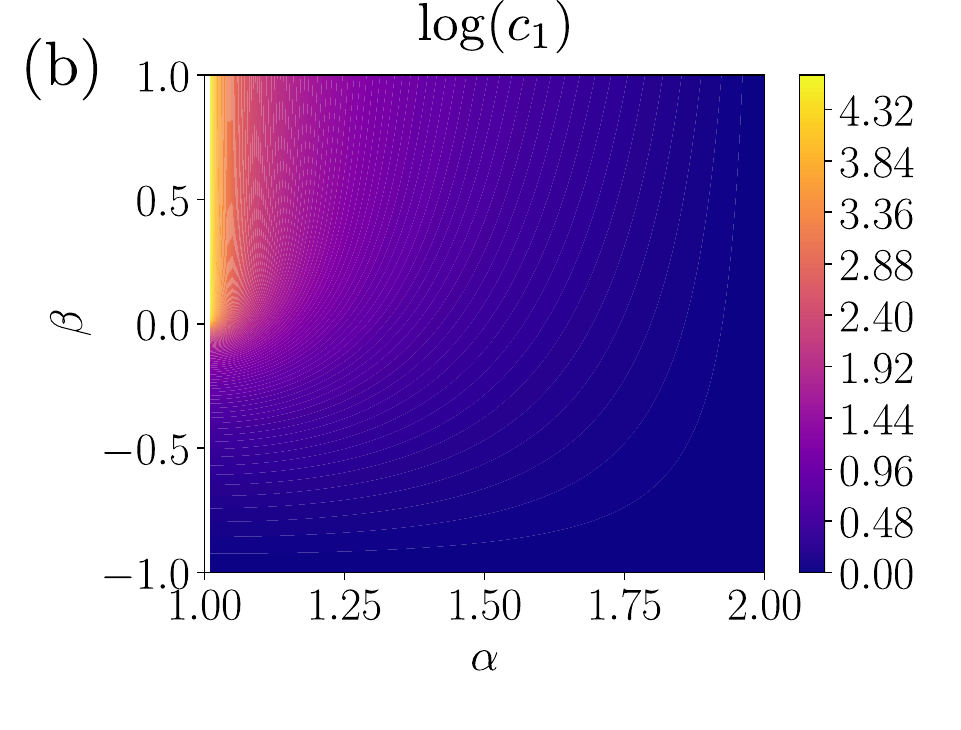}
\caption{The critical exponent $c_1$ given in equation (\ref{eq:cn_fpt_general}) depends strongly on $\alpha,\beta$. Panel a): $c_1$ versus $\beta$ for different values of $\alpha$. Panel b): filled contour plot of $\log(c_1)$ in the $(\alpha,\beta)$ domain. The value of $c_1$ increases without bounds as $\alpha\to 1^+$, and $\beta\to 1^-$.}
\label{fig:cn_vs_al_beta}
\end{figure}

Figure \ref{fig:Xon_Ton_vs_mu}c	 shows that the average value of $X$ during on phases becomes independent of $\mu$ for small $\mu$, like in the Gaussian case. For the mean, which scales as $\langle X\rangle \propto \mu^{c_1}$ at small $\mu$, this implies the following expression for the critical exponent
\begin{equation}
c_1 = \alpha[ m(\alpha,\beta)+2] /(\alpha-1) = \frac{\alpha}{(\alpha-1)} \left( \frac{1}{2} -(\alpha \pi) \arctan(\beta \tan(\alpha \pi/2)) \right). \label{eq:cn_fpt_general}
\end{equation}
The dependence of this result on $\alpha,\beta$ is visualised in figure \ref{fig:cn_vs_al_beta}. The exponent $c_1$ increases monotonically with $\alpha$ and with $\beta$, and is bounded below by $1$. It is equal to unity for all $1<\alpha<2$ when $\beta=-1$.

\begin{figure}[hbt!]
\centering
\includegraphics[width=0.49\textwidth,height=0.4\textwidth]{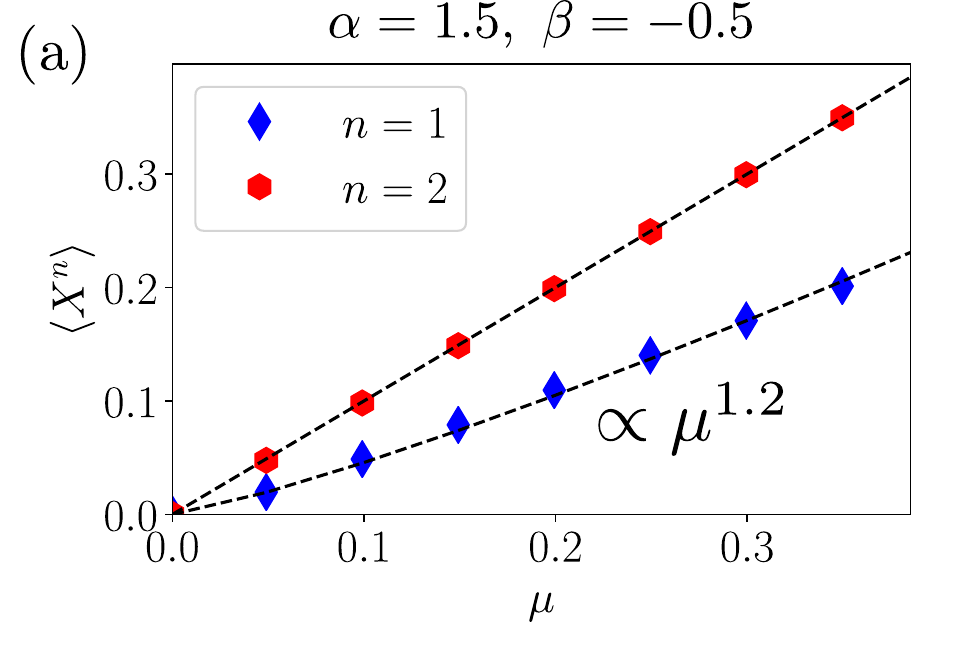}
\includegraphics[width=0.5\textwidth,height=0.4\textwidth]{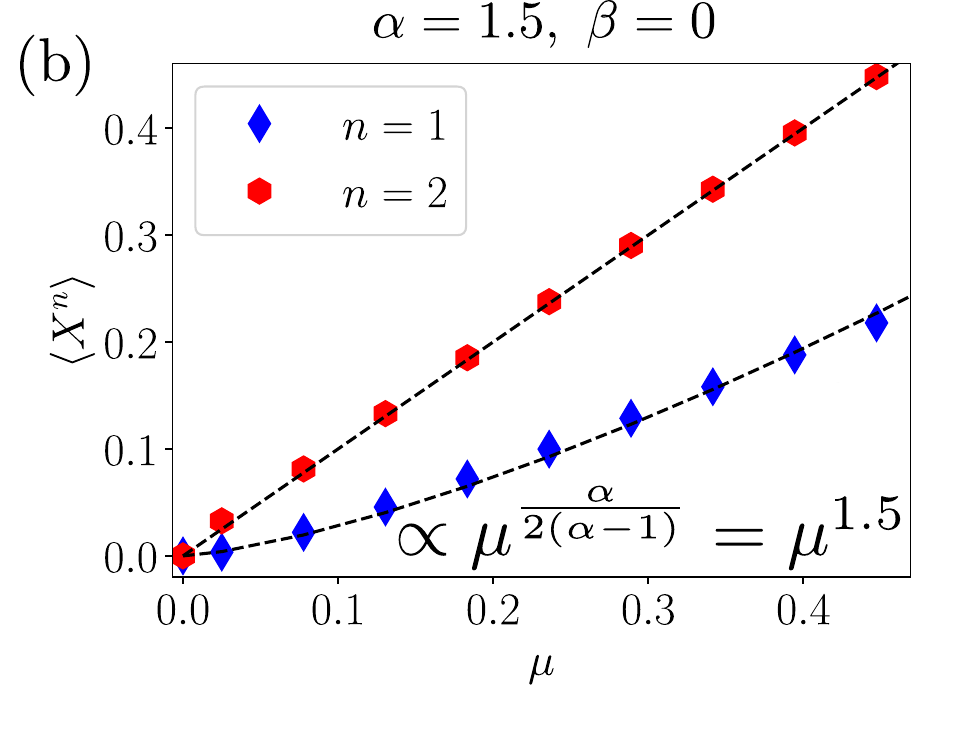}\\
\includegraphics[width=0.5\textwidth,height=0.4\textwidth]{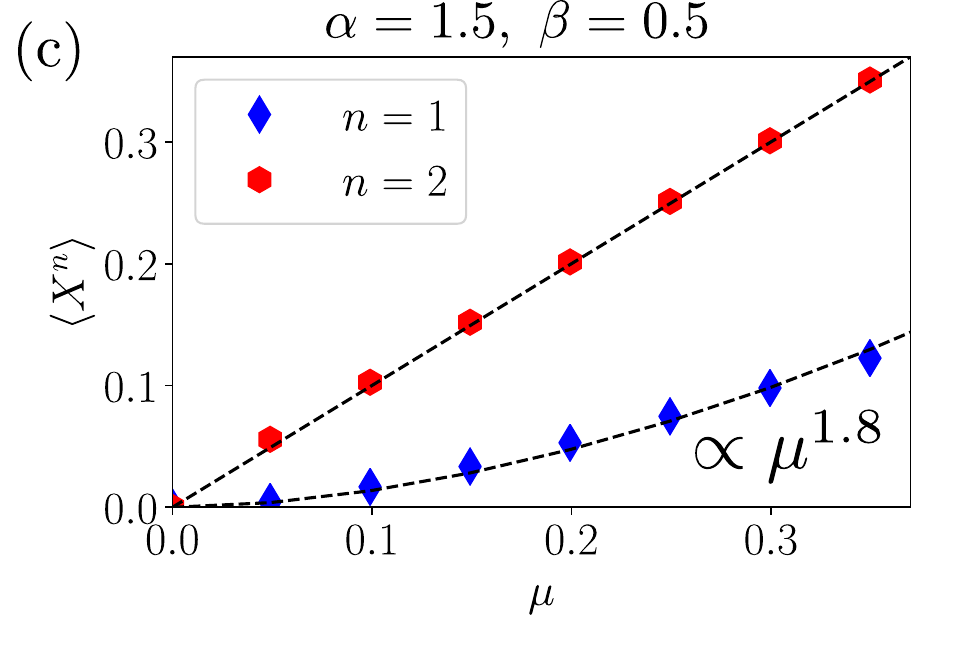}
\caption{Theoretically predicted scaling of moments $\langle X^n\rangle$ compares favourably with numerical solutions of fractional Fokker-Planck equation associated with equation (\ref{eq:langevin}). Symbols show numerical solutions of the fractional Fokker-Planck equation, obtained as described in \cite{vankan2021levy}. Dashed lines indicate the scaling given in by equations (\ref{eq:x2_exact}) and (\ref{eq:cn_fpt_general}). The results shown were computed for $\alpha=1.5$, and $\beta=-0.5$ (panel a), $\beta=0$ (panel b), and $\beta=0.5$ (panel c).} 
\label{fig:compare_critical_scaling}
\end{figure}

The expression simplifies for  $\beta=0$, where $m=-3/2$, and we find spefically
\begin{equation}
c_1=\frac{\alpha}{2(\alpha-1)}. \label{eq:cn_beta0_fpt}
\end{equation}
Equation (\ref{eq:cn_beta0_fpt}) agrees with eq. (\ref{eq:cn_beta0}) for $\nu=1-\frac{\alpha}{2}$; for $\alpha=1.5$, this gives $\nu=0.25$, which is indeed the value found numerically in \cite{vankan2021levy}. For general $\beta$, we can infer by comparison of equations (\ref{eq:cn_beta0}) and (\ref{eq:cn_beta0_fpt}) that
 \begin{equation}
\nu = 1- \alpha [m(\alpha,\beta)+2],
\end{equation}
with $m(\alpha,\beta)$ given in eq. (\ref{eq:m_alpha_beta}). Hence, the expression in eq. (\ref{eq:cn_fpt_general}) improves significantly on the results of \cite{vankan2021levy} by providing the explicit dependence of the critical exponent on the noise parameters $\alpha$ and $\beta$. In figure \ref{fig:compare_critical_scaling}, the prediction of equation (\ref{eq:cn_fpt_general}) is compared to numerical results for $\alpha=1.5$ and $\beta=-0.5,0,0.5$, obtained by integrating the fractional Fokker-Planck equation associated with equation (\ref{eq:langevin}) as described in \cite{vankan2021levy}. The numerical results compare favourably with the predictions. We note, furthermore, that for $\beta=-1$, one finds $m=-1-1/\alpha$ and hence a critical exponent $c_1$ of unity, which is precisely what is found from the stationary PDF in \cite{vankan2021levy}. The reason why we do not directly use time series data generated from (\ref{eq:langevin}) to verify (\ref{eq:cn_fpt_general}), and instead resort to the Fokker-Planck equation, is that the latter approach is more accurate at reduced numerical cost: the PDF can be computed directly, rather than sampling long, highly intermittent time series.

\REV{The main focus of the above discussion is on the non-trivial scaling of the first moment $\langle X\rangle$, since for $-1<\beta<1$, this is the only finite integer moment, apart from $\langle X^2\rangle\propto \mu$, which is always fixed by the exact identity (\ref{eq:x2_exact}), in agreement with the numerical results shown in figure \ref{fig:compare_critical_scaling}. We reiterate that the heuristic argument presented above does not capture the linear scaling of $\langle X^2\rangle$, since the approximation of the on domain as a finite interval breaks down there, due to the logarithmically slow convergence at $X\to \infty$, which requires taking into account contributions from large $X$. More generally, if the cubic nonlinearity is replaced by one of order $s$, then the first $s-2$ integer moments exist. Since the asymptotics of the stationary PDF given in equation (\ref{eq:asymp_pdf_beta_gt_m1}) remain of power-law form up to logarithmic corrections when higher-order nonlinearities are considered, it is reasonable to expect that the scaling exponents derived here for $\langle X\rangle$ would apply to all moments of order $s-2$ and below, but verifying this will require a more detailed investigation, which is left for a future study. Specifically, it would need to be checked that $\langle X^n\rangle_{on}$ becomes independent of $\mu$ as $\mu\to 0$ for $n=1,\dots,s-2$.} \\In summary, the critical exponents predicted here based on L\'evy flight first-passage times are consistent with the results of \cite{vankan2021levy}. Moreover, the present result (\ref{eq:cn_fpt_general}) goes further than \cite{vankan2021levy}, in that it determines the explicit dependence of the critical exponent on $\alpha,\beta$. Hence, the above derivation based on first-passage times, although it may at first sight seem conceptually more complex than the direct computation of moments from the stationary PDF in \cite{vankan2021levy}, provides added value.

\section{Spectral analysis of on-off intermittency}
\label{sec:flicker}
In this section we give a brief summary of $1/f$ noise in on-off intermittency induced by Gaussian noise and present a spectral analysis of Lévy on-off intermittency. We stress again that we use the term $1/f$ noise broadly to refer to low-frequency spectra of power-law form with exponent less than $0$ and greater than $-2$.
\subsection{The Gaussian case}
\begin{figure}
\centering
\includegraphics[width=0.49\textwidth,height=0.398\textwidth]{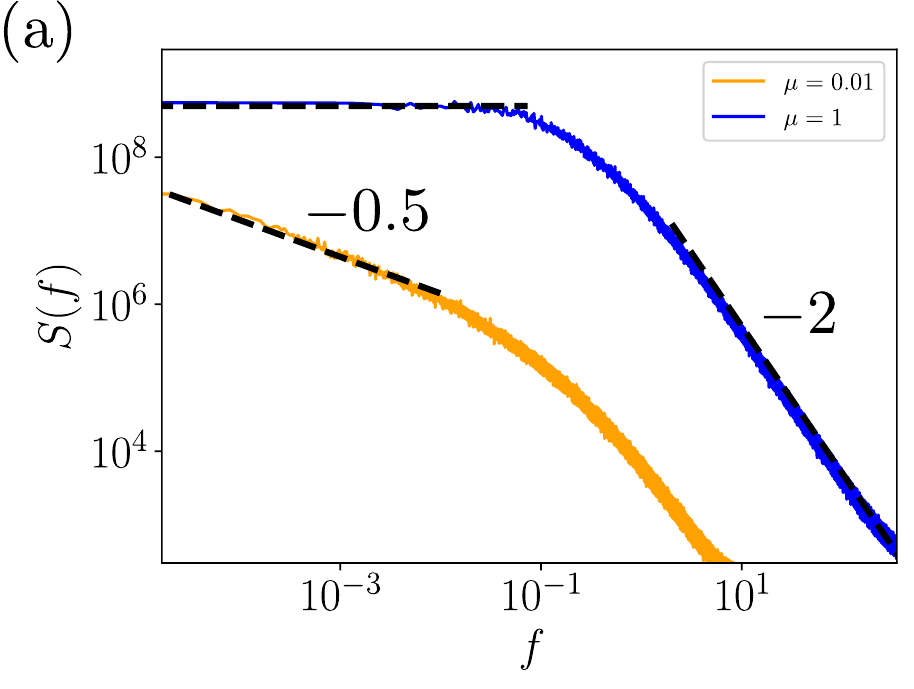}
\includegraphics[width=0.49\textwidth,height=0.39\textwidth]{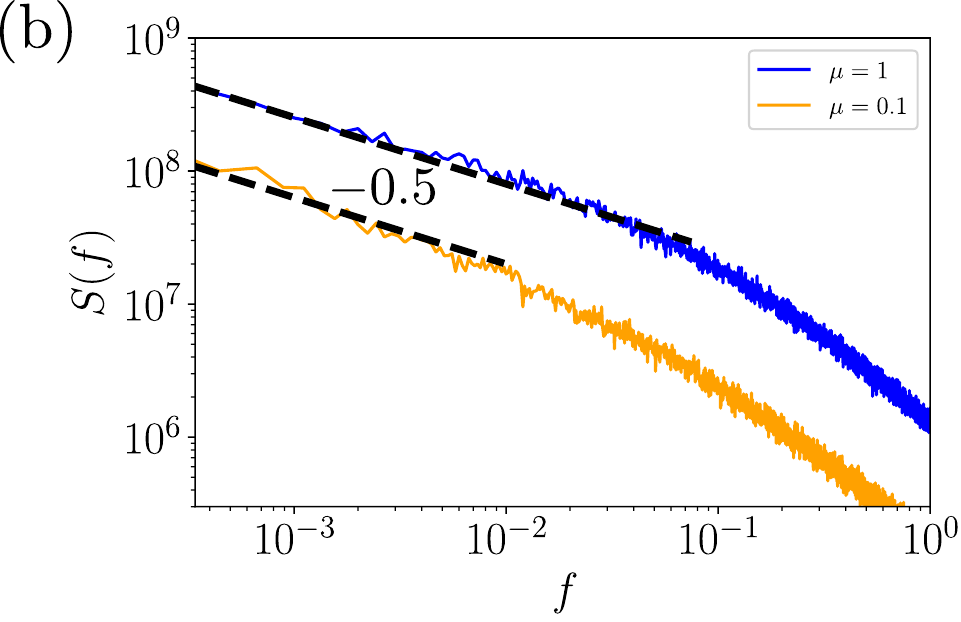}
\caption{Power spectral density $S(f)$ of $X$, computed from time series like that shown in figure \ref{fig:Xon_Ton_vs_mu}a. Panel a): Gaussian noise ($\alpha=2$). The spectra have different shapes for $\mu = 1$ and $\mu=0.01$. At $\mu=1$, the spectrum is close to that of an Ornstein-Uhlenbeck process: it shows a power-law with exponent of approximately $-2$ at high frequencies, and it becomes flat at small $f$. At $\mu=0.01$, a power-law range with exponent $-0.5$ appears at small frequencies, indicative of $1/f^{0.5}$ noise. Panel b) Symmetric Lévy noise with $\alpha=1.5$, $\beta=0$. The shape of the spectrum is qualitatively independent of $\mu$. At low frequencies, there is a power law with exponent $-0.5$, as predicted by (\ref{eq:kappa_m_rel}). }
\label{fig:specs_vs_mu}
\end{figure}

An important known feature of equation (\ref{eq:langevin}) with Gaussian white noise (i.e. Lévy white noise with $\alpha=2$) is that on-off intermittency only occurs within a finite interval of the control parameter $\mu$, where the most probable state is $X=0$. For larger $\mu$, the evolution of $X$ be regarded as (small) fluctuations $X'$ about the mean value $\langle X\rangle$. Heuristically, one can linearise equation (\ref{eq:langevin}) in $X'$, to find that $X'$ approximately obeys an Ornstein-Uhlenbeck process \cite{gardiner1985handbook}, whose power spectrum $S(f)$ is known exactly. It has the property that $S(f)=const.$ at small $f$, i.e. that the low-frequency part of the signal $X(t)$ is white noise, while $S(f)\propto f^{-2}$ at large frequencies. Figure \ref{fig:specs_vs_mu}a shows that this is precisely the form of the power spectrum obtained from a numerical solution of equation (\ref{eq:langevin}) with Gaussian noise at $\mu=1$. By contrast, at $\mu=0.01$ the spectrum features a power-law with exponent $-0.5$ at low frequencies, indicative of $1/f$ noise.
\subsection{A heuristic argument}
It has long been known that intermittency and $1/f$ noise are intimately linked. An insightful early discussion of this topic was given by Manneville in \cite{manneville1980intermittency}. Here we will describe a generalised form of the argument given there, which explains the low-frequency power-law, leveraging our knowledge of the exact asymptotic form of the first-passage time distribution, $p(\tau) \propto \tau^m$. We keep $-2<m<-1$ arbitrary in the argument for the sake of generality. \REV{The reason why the following arguments apply to \textit{low} frequencies is their reliance on \textit{long}-waiting-time asymptotics.}

Consider a long, on-off intermittent time series of total length $T$, generated from eq. (\ref{eq:langevin}). The average time spent in a given off-phase can be computed as $\langle T_{off}\rangle_T \approx \int_0^T p(\tau) \tau d\tau\propto T^{m+2}$, with the broad first-passage time distribution $p(\tau)$. The number of off phases during $T$ is hence $N(T)\approx T/\langle T_{off}\rangle_T\propto T^{-m-1}$. By construction, this is also the number of on phases. Since their average duration is finite, the total time spent in on phases is proportional to $ N(T)$. Hence, the fraction of time spent in the on state is proportional to $T^{-m-1}/T=T^{-m-2}$. This information allows us to estimate the correlation function $C(t) = \langle X(0) X(t)\rangle $, by noting that the only realisations contributing to the ensemble average are those for which $X(t)$ is in an on phase. As argued above, this happens in a fraction of cases that is proportional to $t^{-m-2}$. Thus we obtain $C(t)\propto t^{-m-2}$. The power-law range in the power spectral density (PSD) $S(f)$ of $X$ then follows from the Wiener-Khintchine theorem \cite{wiener1930generalized,khintchine1934korrelationstheorie}, which states that \begin{equation}
S(f)= \int e^{ift} C(t) dt \propto f^{\kappa},\hspace{0.5cm} \textrm{with} \hspace{0.5cm} \kappa= m+1. \label{eq:kappa_m_rel}
\end{equation}
Equation (\ref{eq:kappa_m_rel}) is a general result for bursting signals. It applies, among others, to pressure signals in turbulent fluid flows \cite{abry1994analysis,herault2015low,pereira20191}. Often the power-law exponents $m,\kappa$ are known as $-\alpha$ and $-\beta$, respectively, but here these labels are already used up for the L\'evy noise parameters. For the case of Gaussian noise, one has $m=-1.5$ and thus equation (\ref{eq:kappa_m_rel}) gives $\kappa=-0.5$. This agrees with the numerical results shown in figure \ref{fig:specs_vs_mu}a at small $\mu$, where on-off intermittency occurs.

\begin{figure}[h]
\centering
\includegraphics[width=0.7\textwidth]{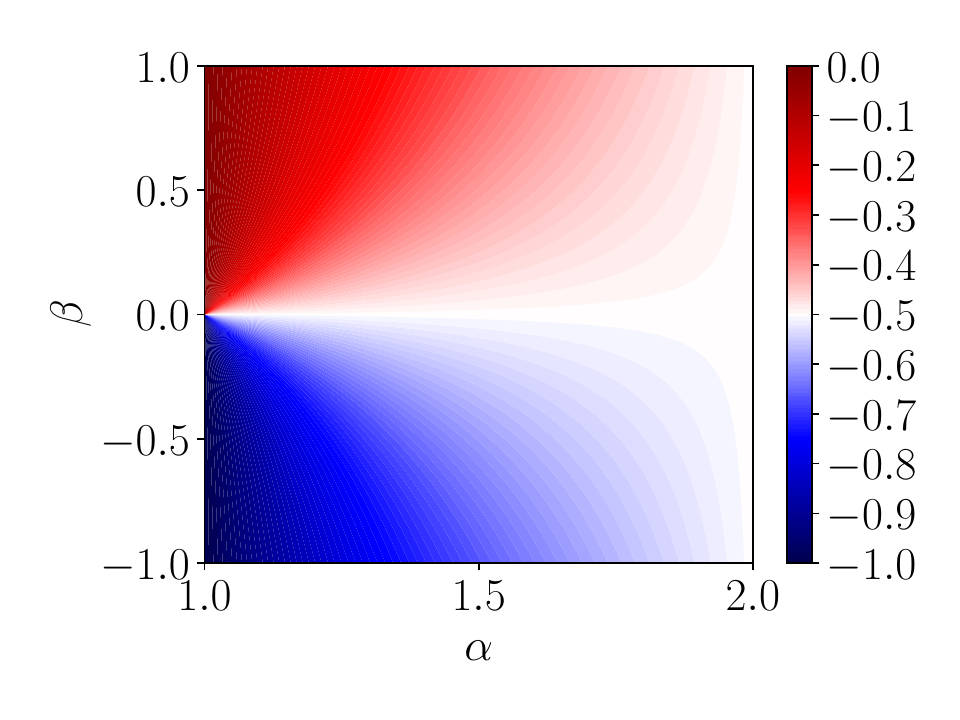}
\caption{Contour plot of the low-frequency spectral exponent $\kappa$, as given in equation (\ref{eq:kappa_m_rel}), in the two-dimensional parameter space spanned by the noise parameters $(\alpha,\beta)$. The value of $\kappa$ is bounded below by $-1$ and bounded above by $0$; $\kappa$ increases monotonically with $\beta$, over a range centered on $-0.5$ which increases  as $\alpha\to 1$. Cf. figure \ref{fig:m_al_beta}.}
\label{fig:kappa_vs_al_beta}
\end{figure}
\begin{figure}
\centering
\includegraphics[width=0.49\textwidth,height=0.381\textwidth]{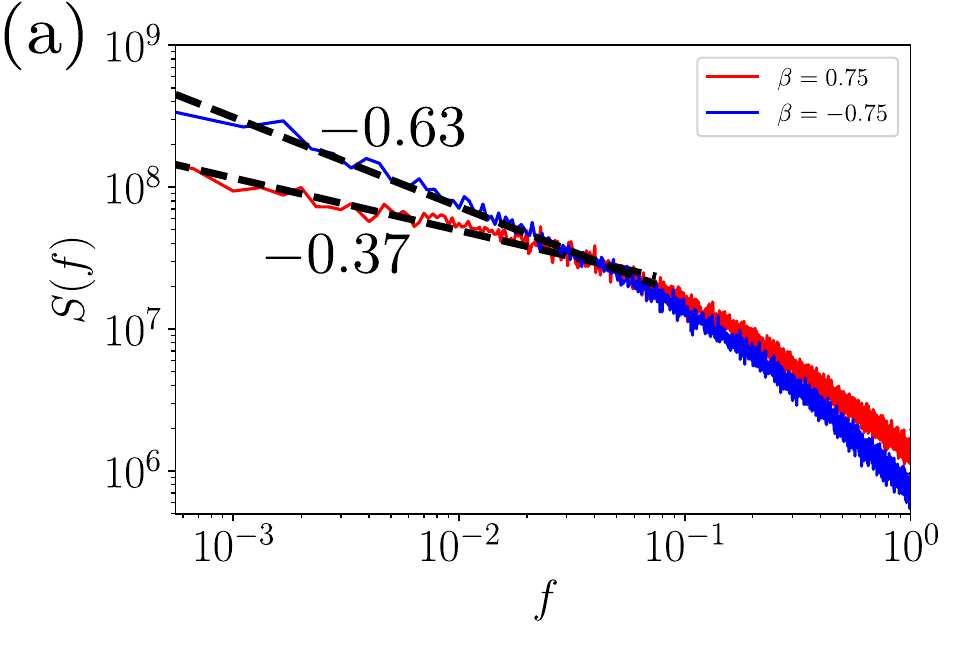}
\includegraphics[width=0.49\textwidth,height=0.38\textwidth]{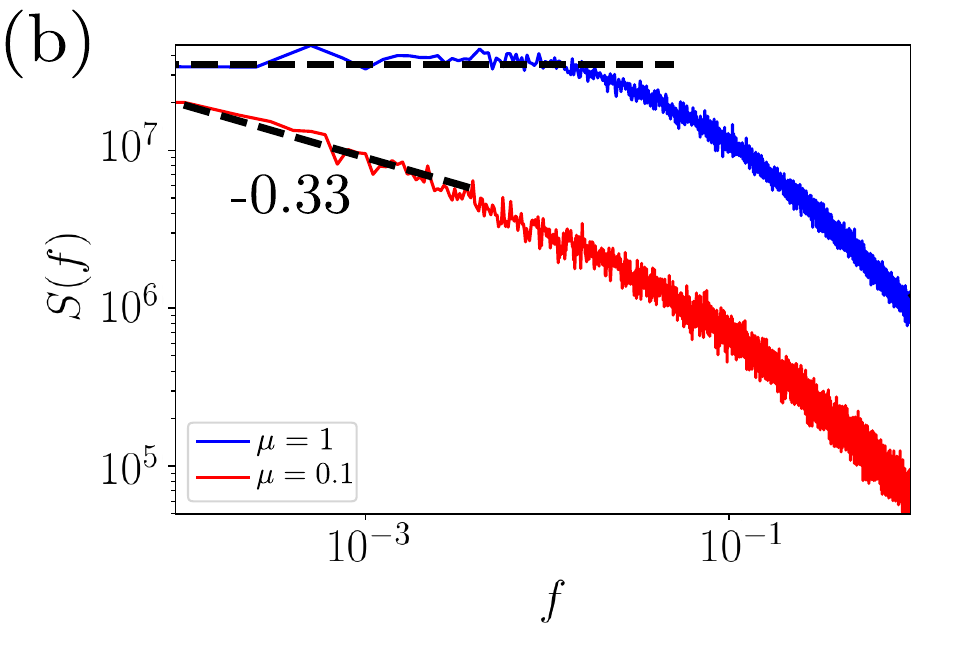}
\caption{Log-log plots of the power spectral density $S(f)$ of $X$, versus frequency $f$, for asymmetric noise with $\alpha=1.5$, at $\mu=1$, $\gamma=1$. Panel a) $\beta=\pm 0.75$. Thick, dashed lines show power laws with the exponent predicted by equation (\ref{eq:kappa_m_rel}). The predicted power laws are compatible with the numerically observed spectra; The low frequency spectrum at $\beta<0$ ($\beta>0$) is steeper (flatter) than in the case of Gaussian noise. Panel b) shows the pecial case $\beta=1$, where on-off intermittency and $1/f$ noise only exist within a finite interval of $\mu>0$ for any $1<\alpha\leq 2$. At $\mu=1$, the spectrum becomes flat at low frequencies, and the high-frequency tail shows an approximate power law $-1.5$. At $\mu=0.1$, by contrast, there is $1/f^{|\kappa|}$ noise (thick dashed line) with an exponent consistent with equation (\ref{eq:kappa_m_rel}).}
\label{fig:ps_asym_noise}
\end{figure}
\subsection{The Lévy case: low frequencies}
Let us now consider the case $1<\alpha<2$, i.e. strictly Lévy on-off intermittency. The dependence of $\kappa$ on $\alpha,\beta$ is shown in figure \ref{fig:kappa_vs_al_beta}. We first focus on the low-frequency part of the spectrum. The spectral exponent $\kappa$ predicted in equation (\ref{eq:kappa_m_rel}) can take any value $\kappa\in(-1,0)$ depending on the choice of $\alpha$ and $\beta$, since $m(\alpha,\beta)\in(-2,-1)$. In particular, for symmetric noise ($\beta=0$), where $m=-1.5$, the low-frequency behaviour of the spectrum $S(f)$ is predicted to be a power law with exponent $-0.5$, independently of $\alpha$.

As discussed in the introduction, Lévy on-off intermittency persists at all values of the control parameter $\mu>0$, for all $1<\alpha<2$ and $\beta<1$. Only in the special case $\beta=1$, it is limited within a finite interval of $\mu$ close to $\mu=0$. Based on this fact, we expect to observe the $1/f$-type noise associated with this on-off intermittency, independently of whether $\mu$ is large or small, except in the special case $\beta=1$. Figure \ref{fig:specs_vs_mu}b confirms this expectation in the case $\alpha=1.5$, $\beta=0$: a spectrum of the form $f^\kappa$ is found both at $\mu=1$, and $\mu=0.1$, with a spectral exponent $\kappa=-0.5$ which is consistent with equation (\ref{eq:kappa_m_rel}).

Figure \ref{fig:ps_asym_noise}a shows the case of asymmetric noise: $\beta=\pm0.75$, at $\alpha=1.5$. At $\beta=-0.75$, the low-frequency spectrum is steeper than in the symmetric case, $\kappa\approx-0.63$, and at $\beta=0.75$, the spectrum is flatter than the symmetric case, $\kappa\approx-0.37$. The values of the observed power-law exponents are in agreement with the theoretical prediction. Finally, the case $\beta=1$ is an interesting singularity in the following sense. As mentioned earlier, on-off intermittency persists at all $\mu>0$, provided that $\beta<1$. At $\beta=1$, however, there is a finite range of $\mu$ where on-off intermittency is observed. This results in the spectra shown in figure \ref{fig:ps_asym_noise}b: at $\beta=1$, the spectrum is flat at low frequencies for $\mu=1$, as in the Gaussian case $\alpha=2$. For $\mu =0.1$, however, there is $1/f^\kappa$ noise with $\kappa\approx-0.33$ in agreement with the prediction of equation (\ref{eq:kappa_m_rel}).

\subsection{The Lévy case: high frequencies}
\begin{figure}
\centering
\includegraphics[width=0.8\textwidth]{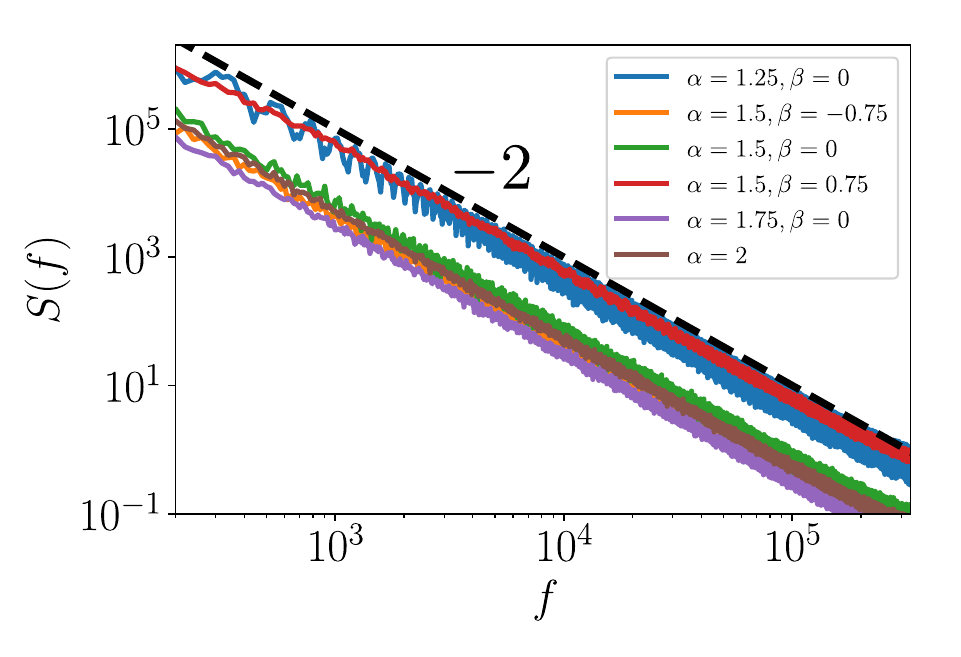}
\caption{High-frequency power spectra $S(f)$ for various $\alpha,\beta$ at $\mu=1,\gamma=1$, generated using the formal solution to (\ref{eq:langevin}) given in \cite{vankan2021levy} with time step $dt=10^{-6}$. In all cases, the high-frequency tail is of power-law form with an exponent close to $-2$.}
\label{fig:hf_specs}
\end{figure}

The results presented so far pertain to the low-frequency range of Lévy on-off intermittency. At high frequencies, the heuristic argument used in the case of Gaussian noise, based on the known spectrum of the Ornstein-Uhlenbeck process, is no longer applicable for Lévy noise, since the Lévy version of the Ornstein-Uhlenbeck process has an infinite second moment, see \cite{chechkin2004levy}, and hence defining a spectrum in terms of the correlation function is not possible. The problem of theoretically computing the \textit{high}-frequency spectrum of Lévy on-off intermittency is therefore more complicated. We numerically calculate $S(f)$  at $f\gg 1$ for different $\alpha,\beta$ by performing simulations with a small timestep $dt=10^{-6}$, averaging over 300 realisations, to obtain the spectra shown in figure \ref{fig:hf_specs}. For all $\alpha,\beta$ we investigated, the high-frequency spectrum has a power-law with exponent close to $-2$. This is consistent with the results obtained by \cite{kharcheva2016spectral}, for the case of additive Lévy noise in a steep potential. There too, the high-frequency power spectrum is found to have an exponent $-1-\omega$, with $\omega$ close to $1$ for all $\alpha\in(1,2)$, although only $\beta=0$ was investigated. \REV{One can anticipate intuitively that the high-frequency behavior is similar for multiplicative and additive noise. This is because the short-time contributions to the correlation function derive from the on phase, where the value of $X$ is large, so that the noise amplitude is constant to leading order. The observed agreement between the result pertaining to additive noise and the present case of multiplicative noise at high frequencies confirms this intuition. By contrast, the non-trivial low-frequency spectral range discussed in the previous section derives from the multiplicative nature of the noise.}

\section{Conclusions}
\label{sec:conclusions}
In this article, we have used exact results on the asymptotic first-passage time distribution of Lévy flights to study anomalous scaling and $1/f$ noise for arbitrary noise parameters $\alpha,\beta$ for Lévy on-off intermittency obeying equation (\ref{eq:langevin}). Both critical exponents and low-frequency spectral power-law exponents were obtained explicitly by heuristic arguments. We have validated the results using numerical solutions of the fractional Fokker-Planck equation associated with equation (\ref{eq:langevin}), as well as direct time integration of the Langevin equation (\ref{eq:langevin}). Moreover, we have shown numerically that the high-frequency power spectrum is of power-law form with an exponent close to the value for Lévy flights in steep symmetric potentials. 

Our results illustrate the non-universality of critical exponents in noisy systems. \REV{In both the Gaussian and Lévy noise cases, the multiplicative nature of the noise causes anomalous scaling, but the scaling exponents are sensitive to the type of noise: for Lévy noise, the solution $X$ of (\ref{eq:langevin}) exhibits $\langle X\rangle \propto \mu^{c_1}$ with a critical exponent which can take any value between $1$ and $+\infty$, depending on the values of the noise parameters $\alpha,\beta$. By contrast, in the case of Gaussian noise, $\langle X^n\rangle\propto \mu$ at small $\mu$, independently of $n$.} In addition to being anomalous (differing from dimensional analysis), the scaling exponents reported here are also an example of multiscaling, since the critical exponent of the second moment $c_2$ is different from $2 c_1$. \REV{Moreover, the $1/f$-type noise generated by Lévy on-off intermittency is of particular interest, since its low-frequency spectral exponent $\kappa$ can be tuned to take any value in $(-1,0)$, depending on $\alpha,\beta$.} This exemplifies that instabilities subject to non-Gaussian noise can display a rich variety of physical behaviours. 

Many directions remain yet to be explored, including the behaviour of the system under truncated Levy noise \cite{schinckus2013physicists}, combined Lévy-Gaussian noise \cite{zan2020stochastic}, finite-velocity Lévy walks \cite{xu2020levy}, different nonlinearities \cite{petrelis2006modification} and higher dimensions \cite{alexakis2021symmetry,graham1982stabilization,alexakis2009planar}. Other problems of interest concern noise with memory, of which few studies exist to date, such as \cite{petrelis2012anomalous,alexakis2012critical}.

\section*{Acknowledgements}
\REV{The authors thank two anonymous referees for their helpful remarks.}
AvK acknowledges partial support from Studienstiftung des deutschen Volkes and the National Science Foundation (grant DMS-2009563). AvK thanks Edgar Knobloch for pointing out useful references.
\section*{References} 
\bibliographystyle{unsrt}	
 \bibliography{lit}

\end{document}